\documentclass[preprint,prl,showpacs,floatfix]{revtex4-1}
\usepackage[final]{graphicx}
\usepackage[usenames]{color}
\usepackage{afterpage}
\usepackage{bm}
\usepackage{multirow}

\date{\today}
\begin{document}

\title{Dielectric screening in two-dimensional insulators: Implications for
excitonic and impurity states in graphane}

\author{Pierluigi Cudazzo$^1$, Ilya V. Tokatly$^{1,2}$ and Angel Rubio$^{1,3}$}

\affiliation{$^1$ Nano-Bio Spectroscopy group and ETSF Scientific Development Centre, 
  Dpto. F\'isica de Materiales, Universidad del Pa\'is Vasco, 
  Centro de F\'isica de Materiales CSIC-UPV/EHU-MPC and Donostia International Physics Center (DIPC), 
  Av. Tolosa 72, E-20018 San Sebasti\'an, Spain\\ 
  $^2$ IKERBASQUE, Basque Foundation for Science, E-48011 Bilbao, Spain\\
  $^3$ Fritz-Haber-Institut der Max-Planck-Gesellschaft, 
  Theory Department, Faradayweg 4-6, D-14195 Berlin-Dahlem, Germany}

\begin{abstract}
For atomic thin layer insulating materials we provide an exact analytic form of the two-dimensional screened potential. In contrast to three-dimensional systems where the macroscopic screening can be described by a static dielectric constant in 2D systems the  macroscopic screening is non local (q-dependent) showing  a logarithmic
divergence for small distances and reaching the unscreened Coulomb potential for
large distances. The cross-over of these two regimes is dictated by 2D layer
polarizability that can be easily computed by standard first-principles
techniques. The present results have strong implications for describing
gap-impurity levels and also exciton binding energies. The simple model derived
here captures the main physical effects and reproduces well, for the case of
graphane, the full many-body GW plus Bethe-Salpeter calculations. As an
additional outcome we show that the impurity hole-doping in graphane leads to
strongly localized states, what hampers applications in electronic devices. In
spite of the inefficient and nonlocal two-dimensional macroscopic screening we
demonstrate that a simple $\mathbf{k}\cdot\mathbf{p}$ approach is capable to describe the electronic and transport properties of confined 2D systems.
\end{abstract} 


\maketitle

\section{Introduction}

The study of two-dimensional (2D) electronic systems is of great fundamental significance in physics. 
Atomic thin layers allow to address the role Coulomb interactions in confined geometries\cite{2d}. 
In this context the synthesis of graphene\cite{novoselov} has triggered a huge amount of work  
 in the understanding and controlling the properties of this
system. In fact due to its unique electronic properties and the low
dimensionality, graphene is considered as one of the most promising materials
for future carbon-based electronics. Nevertheless, the peculiar gapless
ultra-relativistic energy spectrum of graphene\cite{katsnelson,castro} makes the creation of carbon nanodevices
based on p-n junctions highly nontrivial. Therefore, transforming graphene into
a semiconductor with a conventional electron spectrum keeping its two
dimensionality introduced a  challenge that has been a major line of research in the last years. 
Recently an important step towards graphene electronics has been made with the
synthesis of a fully hydrogenated graphene, named graphane\cite{elias} as well as other chemically functionalized 
graphene-like structures\cite{eda,cheng}. Graphane is a wide
band-gap dielectric\cite{sofo,lebegue} and therefore it may become an important part of
nanoelectronic devices as it opens a way to create 2D p-n
junctions\cite{fiori,behnaz}. Similarly, single-atomic layers containing hybridized domains of graphene and h-BN \cite{ajayan}  have been synthesized and follow a completely different and novel (electronic) phenomenology as compared to high-purity or damaged graphene.
Actually, the search for low-dimensional semiconductors is not only focus on graphene and its derivative
compounds but it is also moving towards other layered systems as for example
MoS$_2$, WS$_2$, MoSe$_2$, MoTe$_2$ and BN which can be efficiently dispersed in
common solvents and can be deposited as individual flakes or formed into films\cite{coleman}.
In fact,  MoS$_2$ monolayer has been now synthesized\cite{kis,splendiani}. Contrarily to the
bulk MoS$_2$, it is a direct gap semiconductor with a band gap of 1.8 eV\cite{mak} and
could be used as single-layer transistor\cite{kis}.      
Thus, it is important at this point to make
a deep analysis of many body effects and in particular of the nature and
functional form of the screening in general low dimensional systems (in particular two-dimensional
semiconductors and insulators). In fact, screening effects play
a fundamental role in determining the electron dynamics, the exciton binding
energy and the effective electron-electron and electron-phonon interactions in
the superconducting state. Moreover the screening dictate the optical and
transport properties of 2D devices so that knowing its behavior in low
dimensional systems is fundamental also for practical applications.     

Although in the past the problem of the screening in low dimensional systems has
been extensively investigated in 2D metals\cite{ando} and semiconducting thin
films\cite{keldysh}, to the best of our knowledge it has never been dealt in a strict 2D
dielectric that will be the focus of this letter. In the present work we
provide a strict 2D derivation of the macroscopic screening derived by Keldysh
as a limiting case of a thin film\cite{keldysh}.
We demonstrate that, contrarily to what happens in 3D systems where
the macroscopic screening is mapped in a dielectric constant, in 2D systems the
macroscopic screening is non local so that in the Fourier space it is described
by a $\mathbf{q}$ dependent macroscopic dielectric function. 

Among the various 2D dielectrics, graphane is not only a promising material for nano devices application, but also is very interesting by itself.
In fact, theoretical works based on first principles calculations predict localized spin
states at hydrogen vacancies\cite{ciraci}, demonstrate the existence of unusual strongly
bound charge-transfer excitons\cite{cudazzo} and indicate that doped graphane is probably a
high $T_c$ superconductor\cite{giustino}.
Therefore in the present work we take graphane as the test system to address in
detail the influence of the 2D screening of the Coulomb potential on the
excitonic states and impurity levels. Our findings are general in scope and can be applied to any other 2D 
insulator as the ones described above.

The present work is organized as follows. First, we derive in a simple electrostatic model the exact two-dimensional
screened potential and compare it with its three-dimensional counterpart. We also provide a very simple and pictorial understanding of that potential in terms of the potential created by a one-dimensional charge distribution which length is determined by the two-dimensional layer polarizability. The effect of the specific 2D macroscopic screening is illustrated
addressing the electronic and optical properties of perfect graphane as well as the electronic levels introduced by
hydrogen vacancies in the layer (hole doping). We close the paper with some brief conclusions and perspectives.

\section{Dielectric screening in 2D insulators}

\begin{figure}[ht]
\includegraphics[clip,width=1.0\linewidth]{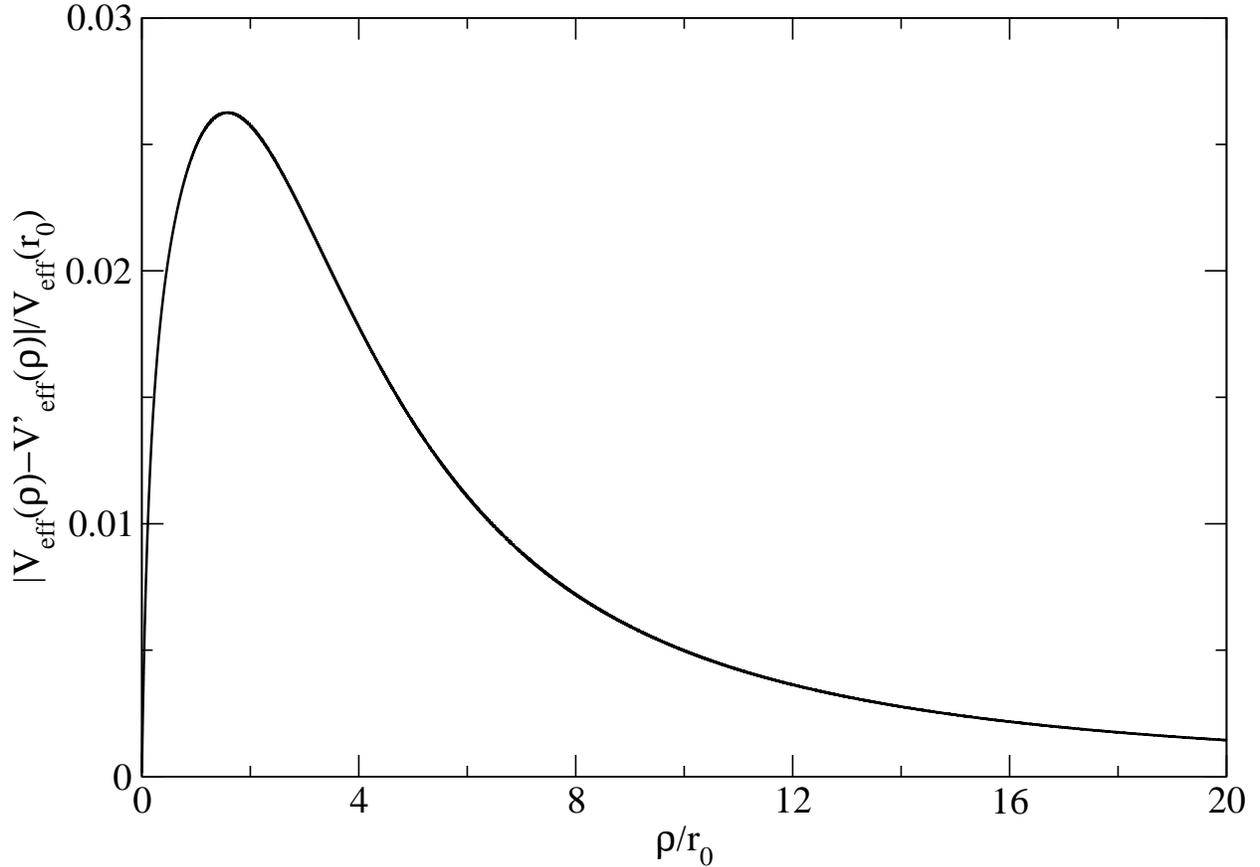}
\caption{Comparison between the true effective potential $V_{eff}(\rho)$ from
Eq.\ref{int2} and its approximated form $V^{\prime}_{eff}(\rho)$ described by
Eq.\ref{int3}.} 
\label{pot}
\end{figure}

To determine a long wavelength static dielectric response of a general 2D insulator we consider a
dielectric sheet of zero thickness at $z=0$ embedded into vacuum, and subject to an external
potential $\phi_{ext}(\mathbf{r})$. For definiteness we assume that
$\phi_{ext}(\mathbf{r})$ is produced by a point charge placed at the origin $n_{ext}({\bf r})=e\delta({\bf r})$. The total electrostatic potential $\phi$ produced by the external
source is related to the total charge density $n$ by Poisson's equation:

\begin{equation}
\nabla^2\phi(\mathbf{r})=-4\pi n(\mathbf{r})
\end{equation}    
where $n=n_{ext}+n_{ind}$ is the sum of the external charge density and
the induced charge density. The induced charge density is confined on the plane $z=0$ and, in the long wave length limit, is related to the 2D macroscopic polarization $\mathbf{P}_{2D}$ ($n_{ind}=-\boldsymbol{\nabla}\cdot\mathbf{P}_{2D}$), which, in turn is proportional to the in-plane component of the total electric field.
Introducing the 2D polarizability $\alpha_{2D}$ of the dielectric sheet, so that
$\mathbf{P}_{2D}(\boldsymbol{\rho})=-\alpha_{2D}\boldsymbol{\nabla}_{\boldsymbol{\rho}}\phi(\boldsymbol{\rho},z=0)$ we obtain an expression of the induced charge density in terms of the macroscopic potential 
evaluated at a point $\mathbf{r}=(\boldsymbol{\rho},z=0)$:
\begin{equation}
n_{ind}(\mathbf{r})=\delta(z)\alpha_{2D}\nabla_{\boldsymbol{\rho}}^2\phi(\boldsymbol{\rho},z=0)
\end{equation}
With this result for the induced charge density the Poisson equation for the
potential of the external point charge takes the form:
\begin{equation}\label{eqpot}
\nabla^2\phi(\mathbf{r})=-4\pi
e\delta(\mathbf{r})-4\pi\alpha_{2D}\nabla_{\boldsymbol{\rho}}^2\phi(\boldsymbol{\rho},z=0)\delta(z)
\end{equation}
while its Fourier transform can be written as:
\begin{equation}\label{poiss}
(|\mathbf{q}|^2+k_z^2)\phi(\mathbf{q},k_z)=4\pi e-4\pi\alpha_{2D}|\mathbf{q}|^2\underbrace{\int\frac{dk_z}{2\pi}\phi(\mathbf{q},k_z)}_{\phi_{2D}(\mathbf{q})}
\end{equation}
where $\mathbf{q}$ is the in-plane component of the wave vector, and the $k_z$-integral in the right hand side defines
the Fourier component $\phi_{2D}(\mathbf{q})$ of the 2D macroscopic
potential. By solving Eq.~(\ref{poiss}) we obtain the following result for $\phi_{2D}(\mathbf{q})$
\begin{equation}\label{int1}
\phi_{2D}(\mathbf{q})=\frac{2\pi e}{|\mathbf{q}|(1+2\pi\alpha_{2D}|\mathbf{q}|)}
\end{equation}
which defines the 2D macroscopic screening of a point charge. As can be seen, for a 2D insulator the macroscopic dielectric screening is no more described by a simple dielectric constant
which renormalizes the electronic charge as in 3D systems. In 2D systems a formally defined dielectric function is intrinsically $\mathbf{q}$-dependent:

\begin{equation}\label{diel}
\epsilon(\mathbf{q})=1+2\pi\alpha_{2D}|\mathbf{q}|
\end{equation}
By the inverse Fourier transform of $e\phi_{2D}(\mathbf{q})$, we can determine
the effective potential $V_{eff}(\mathbf{\rho})$ which is felt by an electron living in the 2D dielectric
in presence of a point charge:

\begin{equation}\label{int2}
V_{eff}(\mathbf{\rho})=\frac{e^2}{4\alpha_{2D}}\left[H_0\left(\frac{\rho}{r_0}\right)
-Y_0\left(\frac{\rho}{r_0}\right)\right]
\end{equation}
where $H_0$ and $Y_0$ are the Struve function and the second kind Bessel
function respectively and $r_0=2\pi\alpha_{2D}$. From the known asymptotic properties of the Struve and Bessel
functions\cite{abramowitz} we determine the following asymptotic behavior of $V_{eff}(\rho)$:

\begin{eqnarray}\label{asi1}
\lim_{\rho\rightarrow\infty}V_{eff}(\rho) & \sim & \frac{1}{\rho} \\ \label{asi2}
\lim_{\rho\rightarrow 0}V_{eff}(\rho) & \sim &
-\frac{1}{r_0}\left[\ln\left(\frac{\rho}{2r_0}\right)+\gamma\right]
\end{eqnarray}
where $\gamma\approx 0.5772$ is the Euler's constant. By the simplest possible
matching of the two asymptotic behaviours we can construct an approximated expression for $V_{eff}(\mathbf{\rho})$ in terms of elementary functions:
\begin{equation}\label{int3}
V^{\prime}_{eff}(\mathbf{\rho})=-\frac{1}{r_0}\left[\ln\left(\frac{\rho}{\rho+r_0}\right)+(\gamma-\ln{2})e^{-\frac{\rho}{r_0}}\right]    
\end{equation}
which gives an accurate description of the effective interaction also at intermediate
values of $\rho/r_0$ as can be inferred from Fig.\ref{pot} 

The above results clearly show that, in contrast to the 3D case, the screening in 2D dielectrics introduces a new length scale $r_0$ which is determined by the the polarizability $\alpha_{2D}$ of the dielectric layer.
When $\rho$ is larger than $r_0$ the effective potential behaves like the 3D
unscreened Coulomb potential while for $\rho\rightarrow 0$ it diverges
logarithmically, i.~e., it goes like the Coulomb potential in two spatial dimensions (the potential of a charged string).
Obviously the logarithmic divergence weakens when $\alpha_{2D}$ increases, 
which means that the screening is more efficient in highly polarizable systems.

\begin{figure}[ht]
\includegraphics[clip,width=1.0\linewidth]{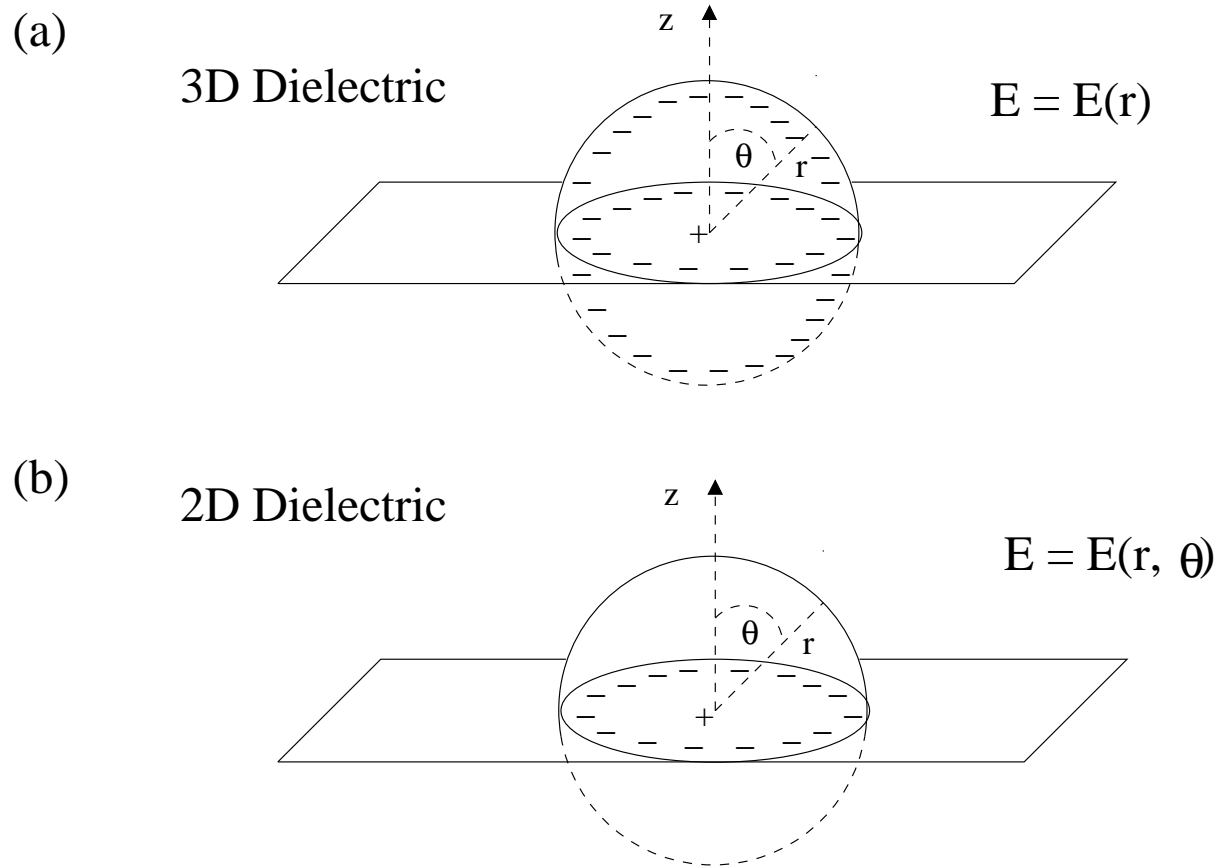}
\caption{Schematic representation of the effect of the macroscopic polarization
induced by a positive point charge on the z=0 plane in 3D (a) and 2D (b) dielectrics.} 
\label{elfield}
\end{figure}

To better understand the difference of the screening in 3D and 2D insulators we
consider a point charge surrounded by a 3D and 2D dielectric medium, respectively
(see Fig.\ref{elfield} (a) and (b)).
The total electric field ($\mathbf{E}$) at a distance $r$ from the point charge
will be the sum of the external field produced by the point charge
($\mathbf{E}_{ext}(\mathbf{r})=\frac{e}{r^2}\hat{\mathbf{r}}$) and the induced field
($\mathbf{E}_{ind}(\mathbf{r})=-4\pi\mathbf{P}(\mathbf{r})$). In 3D dielectrics the latter is equivalent to the electric field produced by a uniform charge distribution on a sphere of radius $r$ centered on the point charge (see
Fig.\ref{elfield} (a)). This charge distribution produces a field of the same functional form as that of the external point charge itself, $\mathbf{E}_{ind}(\mathbf{r})\sim \mathbf{E}_{ext}(\mathbf{r})$, which means that the screening is given by a simple multiplicative renormalization. In the
2D case the situation is quite different. As can be inferred from
Fig.\ref{elfield} (b) , since
the system is polarizable only on the plane, $\mathbf{E}_{ind}$ is
equivalent to the electric field produced by a uniform charge distribution on a
circle of radius $r$. As a consequence it will be a function of
$r$ and $\theta$ with a functional form substantially different from
$\mathbf{E}_{ext}(\mathbf{r})$. This results in a non-local macroscopic screening.

A simple and pictorial understanding of the 2D effective potential $V_{eff}(\rho)$ can be obtained by rewriting Eq.~(\ref{int2}) in a different form. Starting from eq.\ref{int1} we replace the factor $(1+2\pi\alpha_{2D}|\mathbf{q}|)^{-1}$ by its integral representation and rewrite $\phi_{2D}(\mathbf{q})$ as follows
\begin{equation}
\phi_{2D}(\mathbf{q})=\int_{-\infty}^{\infty}dz\frac{2\pi e}{|\mathbf{q}|}e^{-|\mathbf{q}||z|}\frac{e^{-\frac{|z|}{2\pi\alpha_{2D}}}}{4\pi\alpha_{2D}}  .
\end{equation}
Performing the Fourier transform we get the following expression for the
effective interaction between an electron and an external point charge
\begin{equation}
\label{int4}
V_{eff}(\rho)=\int_{\infty}^{\infty}dz\frac{e^2}{\sqrt{\rho^2+z^2}}\frac{e^{-\frac{|z|}{r_0}}}{2r_0}
\end{equation}
Obviously this equation is absolutely equivalent to Eq.~(\ref{int2}), but it is
much more clear physically. Indeed, Eq.~(\ref{int4})
represents the potential (in the plane $z=0$) produced by a one-dimensional charge distribution of the form
\begin{equation}
\label{Q}
Q(\mathbf{r})=e\delta(\boldsymbol{\rho})\frac{e^{-\frac{|z|}{r_0}}}{2r_0}
\end{equation}
Noting that $\int Q(\mathbf{r})d^3{\bf r}=e$, we conclude that in the presence of the dielectric plane the point charge produces a field as it would be effectively smeared out into a 1D string with the charge distribution of Eq.~(\ref{Q}). This behaviour should be contrasted to the multiplicative renormalization of the charge in 3D dielectrics.

Thus, the effect of the 2D dielectric screening can be visualized as follows: two electrons living in a 2D dielectric plane interact as two thin charged rods of the length $\sim 2r_0$ and the line charge density
 $Q(\mathbf{r})\sim e/2r_0$.
The length of the rod sets the characteristic scale of the potential.  From large distances $\rho>>r_0$ the rod is seen as a point charge 
with the potential given by the classical 3D Coulomb law Eq.~(\ref{asi1}). Hence at large distance, the induced polarization is completely inefficient in screening the external field. In the opposite limit $\rho<<r_0$ the rod looks like an infinite wire with the line charge density $\frac{e}{2r_0}$ so that the effective potential reduces to the classical 2D Coulomb potential of Eq.~(\ref{asi2}). Thus at small distance the effect of the induced polarization becomes dominant -- the $1/r$ singularity is replaced by a weaker logarithmic dependence.

It should be noted that our results for the 2D dielectric screening are very closely related to the 
results obtained by Keldysh in ref.~\cite{keldysh} for the interaction potential of two point 
charges in a dielectric slab of the thickness $d$ and characterized by a static
bulk dielectric constant $\epsilon$ (see also Ref.~\onlinecite{jena} and references therein). 
In fact, our Eq.~(\ref{int2}) can be recovered in the limit $\rho\gg d$ and $\epsilon\gg 1$ \cite{keldysh}. The 1D distribution of the effective charge Eq.~(\ref{Q}) can be also viewed as a limiting form of the discrete image charges used to construct the solution of the electrostatic problem for a finite dielectric slab \cite{jena}. The important novel outcome of our derivation is that the form of the effective screened potential of Eq.~(\ref{int2}) is valid even for a microscopically 2D, atomically thin dielectrics for which the notion of the bulk dielectric constant makes no sense. 

The only parameter entering the screened potential of Eqs.(\ref{int1}) and (\ref{int2}) is the polarizability
$\alpha_{2D}$ of the 2D dielectric. Let us show how it can be extracted from the standard ab-initio supercell
calculations where 2D systems are simulated using a periodic stack of layers with
sufficiently large inter-layer distance $L$. For this auxiliary 3D layered system
we can get the 3D macroscopic polarization $P_{3D}=\alpha_{3D}E$, where $\alpha_{3D}$ and $E$
are the 3D polarizability end the total electric field respectively. The macroscopic 3D polarization can be calculated as an
average over $N$ layers in the periodic stack of the microscopic 3D polarization $P^{mic}_{3D}(z)$:

\begin{equation}\label{eq1}
P_{3D}=\frac{1}{NL}\int dzP^{mic}_{3D}(z)
\end{equation}
where, in the definition of $P^{mic}_{3D}(z)$ we have already performed the
one-layer average. Hence $P^{mic}_{3D}(z)$ can be expressed in
terms of the macroscopic 2D polarization $P_{2D}=\alpha_{2D}E_{loc}$ as follows:
\begin{equation}\label{eq2}
P^{mic}_{3D}(z)=\displaystyle\sum_{n=0}^{N}P_{2D}\delta(z-nL)
\end{equation}

with $E_{loc}$ being the local field acting on a single layer.
Inserting Eq.~(\ref{eq2}) in Eq.~(\ref{eq1}) and taking $L$ sufficiently
large so that $E_{loc}\approx E$ we obtain an expression of $\alpha_{2D}$
in terms of $\alpha_{3D}$ (as a generalized Clausius-Mossotti expression for 2D
systems):
\begin{equation}\label{al2d}
\alpha_{2D}=L\alpha_{3D}=L\frac{\epsilon-1}{4\pi}
\end{equation}
where $\epsilon$ is the static dielectric constant of the 3D layered system.  The value of $\epsilon$ entering Eq.~(\ref{al2d}) can be evaluated directly from the first principles calculation of the dielectric function
$\epsilon_{\mathbf{G}\mathbf{G}^{\prime}}(\mathbf{q},\omega)$ as:
\begin{equation}
\epsilon=\lim_{\mathbf{q}\rightarrow 0}\frac{1}{[\epsilon^{-1}(\mathbf{q},\omega=0)]_{\mathbf{G}=\mathbf{G}^{\prime}=0}} .
\end{equation}
When $L$ goes to infinity, $\epsilon$ of the layered system approaches the vacuum dielectric constant, $\epsilon=1+O(1/L)$.
Therefore the $L\to\infty$ limit of the right hand side in Eq.~(\ref{al2d}) yields a finite value that is equal to the 2D polarizability.  In practice one performs calculations for several sufficiently large $L$ to ensure the convergence of $\alpha_{2D}$.

\section{Excitonic and impurity states in graphane}

We apply the results of the previous section to the description of excitonic and impurity states in 
graphane. The problem of excitons in graphane has been addressed recently using a fully {\it ab initio}  many-body 
self-energy GW-BSE approach \cite{cudazzo}. In this section we show that the effective screened potential of Eq.~(\ref{int2}) combined with the $\mathbf{k}\cdot\mathbf{p}$ description of the electronic and hole states leads to a very simple and accurate description of strongly 
bound electron-hole and hole-impurity states as obtained from the GW-BSE calculations.

Graphane is a representative of wide band gap 2D dielectrics, which is
obtained from the ideal graphene by depositing hydrogen atoms on both sides of
graphene plane. The resulting electronic structure is dictated by sp$^3$ hybridization
of the carbon orbitals, which causes the opening of a wide band gap (of 5.4 eV at the $\Gamma$
point) \cite{lebegue}. States at the top of the
valence band belong to E$_g$ 2D irreducible representation of the graphane point
group D$_{3d}$, while the bottom of the conduction band belong to the A$_{2u}$
1D irreducible representation. As shown in Ref.\cite{cudazzo} transitions
from the top of the valence band to the bottom of the conduction band are allowed
in the dipole approximation and result in strong excitonic effects in the absorption spectra. In particular,
the corresponding electron-hole pairs give rise to two nearly degenerate
excitons with binding energy of about 1.6 eV. This large binding energy (one
order of magnitude larger then in typical semiconductors) seems to be surprising
since both valence and conduction bands form almost perfect parabolas in a wide energy-momentum range around the $\Gamma$
point and therefore the excitonic states are expected to be well described in terms of the
effective mass approximation, in spite of their small radius. We will demonstrate explicitly that
the effective mass approximation does indeed works perfectly, and that the unusually
large binding energy is completely explained by the weak and nonlocal 2D screening
discussed in the previous section.

\begin{table}[hc]
\begin{center}
\begin{tabular}{|c|c|c|c|}
\hline
 Angular & Quantum & Exciton & Impurity  \\
momentum ($l$) & number ($n$) & energy (eV) & level (eV) \\
\hline
\hline
\multirow{2}{*}{$l=\pm 1$} & $n=1$ & -1.77 & -2.12 \\
\cline{2-4}
& $n=2$ & -0.67 & -0.90 \\
\hline
\multirow{4}{*}{$l=0$} & \multirow{2}{*}{$n=1$} & $E_+=-1.13$  & $E_+=-1.47$ \\
\cline{3-4}
& & $E_-=-0.78$ & $E_-=-0.92$ \\
\cline{2-4}
 & \multirow{2}{*}{$n=2$} & $E_+=-0.52$  & $E_+=-0.74$ \\
\cline{3-4}
& & $E_-=-0.34$ & $E_-=-0.45$ \\ 
\hline
\multirow{2}{*}{$l=\pm 2$} & $n=1$ & -0.92 & -1.17  \\
\cline{2-4}
& $n=2$ & -0.43 & -0.58 \\
\hline	  
\end{tabular}
\end{center}
\caption{Exciton binding energy and impurity levels in the effective mass
approximation for some selected values of the quantum numbers $n$ and $l$.}
\label{tab}
\end{table} 

Let us start with the $\mathbf{k}\cdot\mathbf{p}$ effective mass approximation for the electronic states in graphane \cite{cudazzo,ilya}. The Hamiltonian for the conduction band is trivially given by 
$\hat{H}_c(\hat{\bf p})=\frac{\hat{p}^2_x+\hat{p}^2_y}{2m_e}$, while for the
valence band Hamiltonian $\hat{H}_v(\hat{\bf p})$ we adopt the representation obtained in Ref.\cite{ilya}:
\begin{equation}
\hat{H}_v(\hat{\bf p})=\frac{1}{2}\alpha I\hat{\bf p}^2  +\frac{1}{4}\beta [\sigma_{+}\hat{p}_{+}^2+\sigma_{-}\hat{p}_{-}^2]
\end{equation}
where $\hat{\bf p}=-i\nabla$ is the in-plane momentum operator,
$\hat{p}_{\pm}=\hat{p}_{x}\pm i\hat{p}_{y}$, $I$ is the identity matrix, and
$\sigma_{\pm}=\sigma_{x}\pm i\sigma_{y}$ with $\sigma_j$ being the Pauli
matrices, and $\alpha=2.62/m_0$, $\beta=0.98/m_0$ and $m_e=0.83m_0$ are the band parameters expressed in terms of the bare electronic mass $m_0$ and obtained from the ab-initio band structures \cite{cudazzo}.

The excitonic Hamiltonian for the zero momentum excitons can then be constructed in standard way \cite{bassani}:
\begin{equation}
\hat{H}_{ex}=\hat{H}_c(\hat{\bf p})+\hat{H}_v(\hat{\bf p})-V_{eff}(\rho)
\end{equation}
where $V_{eff}(\rho)$ is the effective 2D screened electron-hole interaction given by Eq.~(\ref{int2}). Explicitly the final effective mass equation for the relative motion of the electron and the hole takes the following form
\begin{equation}\label{em1}
\left(\frac{1}{2}\gamma_1\hat{I}\hat{\mathbf{p}}^2+\frac{1}{4}\gamma_2[\hat{\sigma}_+\hat{p}_+^2+\hat{\sigma}_-\hat{p}_-^2]
- V_{eff}(\rho)\right)\hat{\Phi}(\boldsymbol{\rho})=E\hat{\Phi}(\boldsymbol{\rho})
\end{equation}
where $\gamma_1=\alpha+\frac{1}{2m_e}$ and $\gamma_2=\beta$.

To classify the eigenstates of Eq.~(\ref{em1}) we note that the Hamiltonian $\hat{H}_{ex}$ commutes with an operator $\hat{L}_z$ that is defined as follows 
\begin{eqnarray}
\label{Lz}
\hat{L}_z = (\boldsymbol{\rho}\times\hat{\mathbf{p}})_z -\sigma_z \equiv (x\hat{p}_y-y\hat{p}_x) -\sigma_z.
\end{eqnarray}
Obviously, the operator $\hat{L}_z$ corresponds to the $z$-component of the total angular momentum, with the second term in Eq.~(\ref{Lz}) being related to the orbital momentum of the local currents inside the unit cell of graphane \cite{ilya}.
Since $[\hat{H}_{ex},\hat{L}_z]=0$ the excitonic states can be classified by the
eigenstates of the total angular momentum operator. In other words, the eigenfunctions of Eq.~(\ref{em1}) can be written in terms
of radial wave functions ($\mathcal{Z}_l(\rho)$, $\chi_{l}(\rho)$) ordered by the
integer quantum number $l$ defining the eigenvalue of $\hat{L}_z$:

\begin{equation}\label{eq3}
\hat{\Phi}_l(\rho,\theta)=\left(
\begin{array}{c}
 e^{i\theta}\mathcal{Z}_l(\rho) \\
 e^{-i\theta}\chi_{l}(\rho) \\
\end{array}
\right)e^{il\theta}
\end{equation}

Inserting the expression of Eq.~(\ref{eq3}) into Eq.~(\ref{em1}) we obtain the equation for the
radial part of the envelop wave functions:

\begin{eqnarray}\label{em2a}
&&\left(-\frac{\gamma_1}{2}\left[\partial^2_{\rho}+\frac{1}{\rho}\partial_{\rho}-\frac{(1+l)^2}{\rho^2}\right]-V_{eff}(\rho)\right)\mathcal{Z}_l(\rho)-
\nonumber \\
&&-\frac{\gamma_2}{2}\left(\partial_{\rho}-\frac{l}{\rho}\right)\left(\partial_{\rho}+\frac{1-l}{\rho}\right)\chi_{l}(\rho)=E_l\mathcal{Z}_l(\rho)
\\
&&\left(-\frac{\gamma_1}{2}\left[\partial^2_{\rho}+\frac{1}{\rho}\partial_{\rho}-\frac{(1-l)^2}{\rho^2}\right]-V_{eff}(\rho)\right)\chi_{l}(\rho)-
\nonumber \\ \label{em2b}
&&-\frac{\gamma_2}{2}\left(\partial_{\rho}+\frac{l}{\rho}\right)\left(\partial_{\rho}+\frac{1+l}{\rho}\right)\mathcal{Z}_l(\rho)=E_l\chi_{l}(\rho)
\end{eqnarray}
Thus, each excitonic state is completely defined by the quantum number $l$ and
the positive integer $n$ denoting the discrete eigenvalues of Eqs.(\ref{em2a}) and (\ref{em2b}) for given $l$. The corresponding microscopic wave function of the exciton for a fixed position $\mathbf{r}_h$ of the hole can be written as follows:
\begin{eqnarray}
\Psi^l_{ex}(\mathbf{r},\mathbf{r}_h) &=&
\mathcal{Z}_l(\boldsymbol{\rho})\psi_e(\mathbf{r})\psi^{(1)}_h(\mathbf{r}_h)+
\nonumber \\ \label{Psi-ex}
&+& \chi_l(\boldsymbol{\rho})\psi_e(\mathbf{r})\psi^{(2)}_h(\mathbf{r}_h)
\end{eqnarray}
where $\psi_e$ is the electron Bloch wave function and $\psi^{(1,2)}_h$ the hole
Bloch wave functions related to the two fold degenerate valence bands.

Analyzing the structure of Eqs.~(\ref{em2a}) and (\ref{em2b}), we observe that for all $l\neq 0$ the system of differential
equations is invariant under the transformation $l\to-l$, $\mathcal{Z}\to\chi$,
$\chi\to\mathcal{Z}$. Therefore all excitonic states with $l\neq 0$ are double
degenerate with $E_l=E_{-l}$, which is a clear consequence of the time-reversal
invariance of the Hamiltonian. We also note that only excitons corresponding to $l=\pm 1$ are dipole active.

The only non-degenerate state corresponds to a dark exciton with zero angular momentum, $l=0$. Interestingly, for $l=0$ the diagonal and off-diagonal operators in the system of Eqs.~(\ref{em2a}), (\ref{em2b}) are equal to each other. As a results the problem reduces to completely decoupled equations for the ``symmetric'' and ``antisymmetric'' states 
\begin{equation}
\left[-\frac{1}{2}(\gamma_1\pm\gamma_2)\left(\partial^2_{\rho}+\frac{1}{\rho}\partial_{\rho}-\frac{1}{\rho^2}\right)-V_{eff}(\rho)\right]\phi_{\pm}(\rho)=E_{\pm}\phi_{\pm}(\rho)
\end{equation}
and the excitonic spinor wave function for $l=0$ takes the form
\begin{equation}
\hat{\Phi}^{\pm}_{l=0}=\frac{1}{\sqrt{2}}\left(
\begin{array}{c}
e^{i\theta}\phi_{\pm} \\
e^{-i\theta}\pm\phi_{\pm} \\
\end{array}
\right)
\end{equation}

To practically solve Eqs.~(\ref{em2a}) and (\ref{em2b}) we expanded the radial part of the envelop wave function
on the 2D hydrogen eigenfunctions $u_{n,l}$\cite{yang}, so that
$\mathcal{Z}_{l}(\rho)=\sum_n a_nu_{n,l+1}(\rho)$ and $\chi_l(\rho)=\sum_n b_nu_{n,l-1}(\rho)$. This complete
orthonormal basis set assures the correct asymptotic behavior of the
eigenfunctions of the excitonic Hamiltonian.

\begin{figure}[ht]
\includegraphics[clip,width=1.0\linewidth]{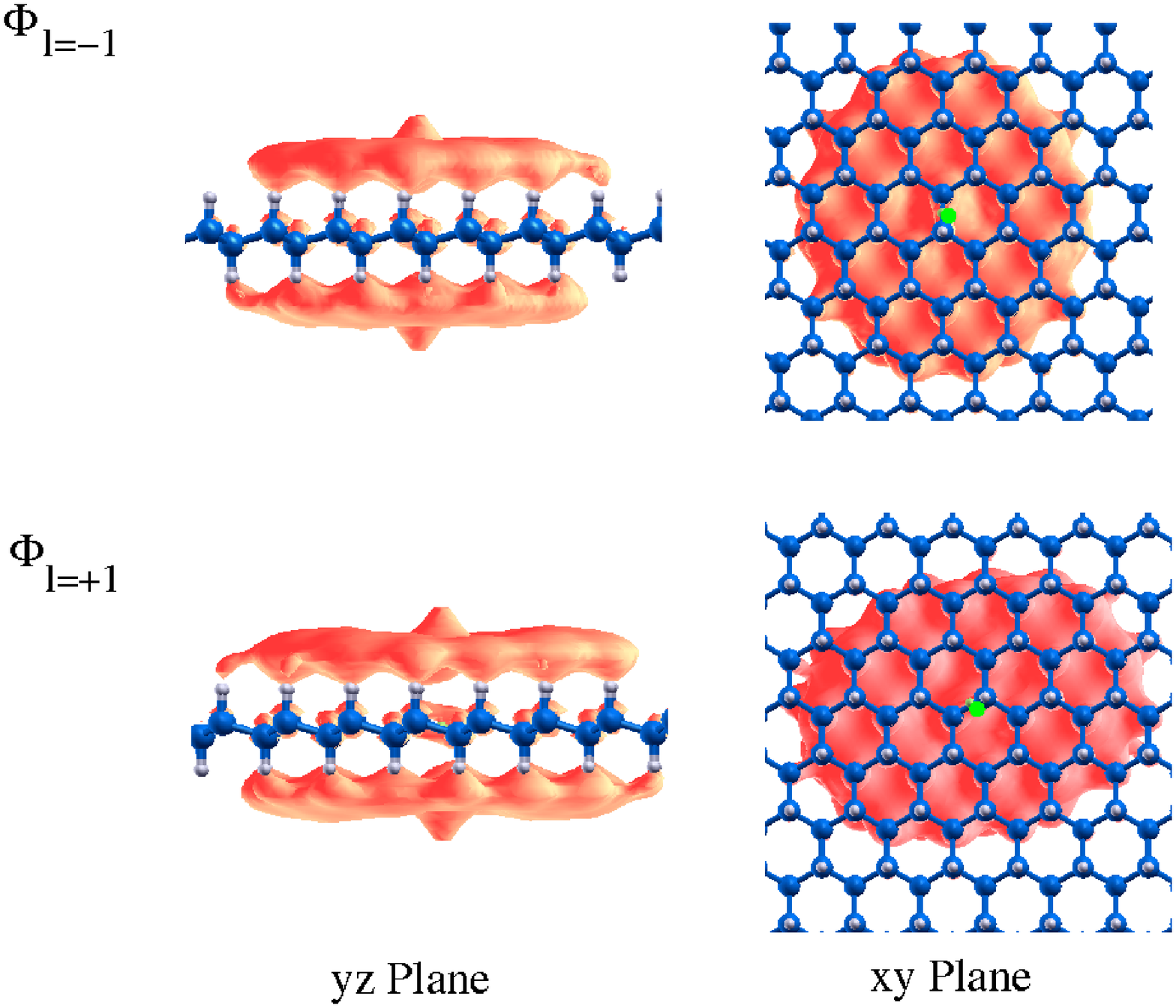}
\caption{3D-Shape of the low energy excitonic wave functions for a fixed position of
the hole (marked as a green circle) as obtained from Eq. (\ref{Psi-ex}). Note
that the shape of the excitonic wave functions is in perfect agreement with that
obtained by the full solution of the BS equation in Ref.\cite{cudazzo}} 
\label{exciton}
\end{figure}

Our results for a selected set of lowest energy states are summarized in table \ref{tab}. As we can see, the ground state of the
excitonic Hamiltonian corresponds to $l=\pm 1$. This state is two-fold degenerate and
optically active with the binding energy $E_{l=\pm 1}=1.77$ eV, which is in a perfect agreement
with the values obtained by solution of the Bethe-Salpeter equation\cite{cudazzo}. The
corresponding excitonic wave functions (Fig.\ref{exciton}), calculated using Eq.~(\ref{Psi-ex}), shows that both excitons are
strongly localized with an average radius of about $11.5$ a.u. As can be
inferred from Fig.\ref{exciton} these excitations give rise to a charge transfer from the
carbon plane towards the hydrogen plane. The first excited state correspond to zero angular momentum
$l=0$. This dark exciton is also found from the solution of the Bethe Salpeter
equation, which is additional confirmation of the present simple theory. 

Therefore our results demonstrate that, despite the large binding energy,
excitons in graphane are indeed described in terms of the effective mass approximation, 
provided the correct form of the effective electron-hole interaction is used (as derived in the present work, 
Eqs.~(\ref{int2}) and (\ref{int4}). We clearly see that the unusual, large binding energy is related to a weak and
nonlocal 2D dielectric screening which is completely
inefficient at large distances. The small overestimation of the exciton binding
energy respect to the ab-initio value may be ascribed to the lack in our
approach of short range contributions to the induced polarization and exchange electron-hole interaction. All these effects can only reduce the exciton
binding energy. As a matter of fact, the effects of short range corrections are small and, if necessarily, can be easily included perturbatively.

Using the same formalism we can now look at the effect of the 2D screening on 
impurity states. First, we focus on the acceptor states in the hole doped graphane, as it is expected to
be the most natural way to dope this system. Indeed naively one may assume that
extra holes are easily introduced by dehydrogenation. In this context it is worth noting that hole doped graphane has been predicted to be a high
T$_c$ superconductor \cite{giustino}. 

Similarly to the excitonic case, the parabolicity of the valence bands in a wide energy range suggests that holes in the presence of hydrogen vacancies can be well described in terms of the effective mass approximation. Therefore acceptor impurity
levels can be obtained by solving Eqs.~(\ref{em2a}), (\ref{em2b})  with $\gamma_1=\alpha$. The corresponding results are presented in table~\ref{tab}. The ground impurity state corresponds to $n=1$ and $l=\pm 1$ and
is characterized by a binding energy of about 2.12 eV in good agreement with the ab-initio value (1.86 eV). This quantity represents
the position of the impurity level with respect to the top of the valence band.
Therefore, for impurity levels the 2D nonlocal screening results in unusual
large binding energy that exceeds by two or three orders of magnitude the corresponding values
for typical semiconductors. Comparing the values of the binding energy with the
gap energy (5.4 eV) we find that the impurity level is close to the center of
the graphane gap. 

Finally, when the dopant is a donor, the electron in the conduction band is
described by a simple 2D hydrogen like Schr\"odinger equation with $V_{eff}(\rho)$ of Eq.~(\ref{int2})
replacing the Coulomb potential. In this case for the lowest bound state 
(corresponding to $n=1$ and $l=0$) we get a binding energy of about 3.15 eV. Therefore for electron doped graphane the specific 2D screening of the impurity potential also causes the formation of mid-gap impurity levels.

The above results lead us to an unfortunate but important conclusion. 
The standard for 3D semiconductors impurity doping, both donor and acceptor, 
will probably not work for graphane and most likely for other atomically thin dielectric. 
In particular a slightly dehydrogenated graphane cannot be considered as a semiconductor 
with extra highly mobile holes in the valence band. All holes will be strongly localized on the hydrogen vacancies with the radius of the bound state of the order of the interatomic distance. The reason for this behavior is a very weak and inefficient screening in 2D dielectric materials. 

\section{Conclusions}

In conclusion, we derived an expression of the macroscopic screening in 2D
dielectrics showing that, contrarily to what happens in 3D systems where the
macroscopic screening is mapped in a dielectric constant, in 2D systems the
macroscopic screening is non local. As a result the effective potential produced
by an external point charge surrounded by a 2D dielectric has a
functional form which is substantially different from the bare Coulomb potential. It presents a logarithmic
divergence for $\rho\rightarrow 0$ and reduces to the unscreened Coulomb
potential at large distances. The 2D polarizability $\alpha_{2D}$ determines the characteristic length scale $r_0$ at which the two asymptotic forms are matched. This behavior strongly modify the optical and
transport properties of 2D systems. In particular
we show that  hole impurity doping leads to strongly bound localized states with low mobility.
Moreover, spite of the inefficient and two-dimensional macroscopic $\mathbf{q}$-dependent screening the simple $\mathbf{k}\cdot\mathbf{p}$ approach works very well to describe the electronic properties up to very high energy, and very short spatial scales. Our results imply that the $\mathbf{k}\cdot\mathbf{p}$ theory supplemented with a proper macroscopic treatment of the 2D screening forms a solid basis for a quantitative description of various, both equilibrium and nonequilibrium, in particular, transport properties of nanostructured 2D systems.

\section*{Acknowledgments}
We acknowledge funding by the Spanish MICINN (FIS2010-21282-C02-01), ACI-promociona project (ACI2009-1036), ``Grupos Consolidados UPV/EHU 
del Gobierno Vasco'' (IT-319-07), and the European Community through e-I3 ETSF project (Contract No. 211956) and THEMA (Contract number: 228539).

\end{document}